\documentclass[a4paper,11pt]{article}
\usepackage{pos}
\usepackage{physics}

\title{IceCube constraints on Violation of Equivalence Principle}

\author*[a,b]{Damiano F. G. Fiorillo}

\affiliation[a]{Dipartimento di Fisica ”Ettore Pancini”, Università degli studi di Napoli ”Federico II”,\\ Complesso Univ. Monte S. Angelo, I-80126 Napoli, Italy}

\affiliation[b]{INFN - Sezione di Napoli,\\ Complesso Univ. Monte S. Angelo, I-80126 Napoli, Italy}

\emailAdd{dfgfiorillo@na.infn.it}

\abstract{Among the information provided by high energy neutrinos, a promising possibility is to analyze the effects of a Violation of Equivalence Principle (VEP) on neutrino oscillations. We analyze the IceCube data on atmospheric neutrino fluxes under the assumption of a VEP and obtain updated constraints on the parameter space with the benchmark choice that neutrinos with different masses couple with different strengths to the gravitational field. In this case we find that the VEP parameters times the local gravitational potential at Earth can be constrained at the level of $10^{-27}$. We show that the constraints from atmospheric neutrinos strongly depend on the assumption that the neutrino eigenstates interacting diagonally with the gravitational field coincide with the mass eigenstates, which is not a priori justified: this is particularly clear in the case that the basis of diagonal gravitational interaction coincide with the flavor basis, which cannot be constrained by the observation of atmospheric neutrinos. Finally, we quantitatively study the effect of a VEP on the flavor composition of the astrophysical neutrinos, stressing again the interplay with the basis in which the VEP is diagonal: we find that for some choices of such basis the flavor ratio measured by IceCube can significantly change.}

\FullConference{37$^{\rm{th}}$ International Cosmic Ray Conference (ICRC 2021)\\
		July 12th -- 23rd, 2021\\
		Online -- Berlin, Germany}

\begin{document}
\maketitle

\section{Introduction}

Despite the enormous successes of the general theory of relativity, a complete theory of gravity consistent with quantum field theory has yet to be constructed. For this reason, testing the underlying principles of general relativity can be a powerful source of information. One of these founding principles is the equivalence principle, which implies that all bodies couple to gravity with the same strength. A large variety of experimental setups are devoted to testing the equivalence principle, including torsion-balance experiments, free-fall experiments, and spectroscopy of atomic levels.

A complementary method of investigation is to look for violations of the equivalence principle (VEP) in the neutrino sector. As well-known, neutrinos come in three different flavor eigenstates. If three different combinations of these flavor eigenstates coupled with different strengths to an external gravitational potential, a relative dephasing would be induced leading to non-standard oscillations. The non-observation of these non-standard oscillations can be used to constrain the magnitude of a possible VEP in the neutrino sector. The frequency of the VEP-induced oscillations in a weak external potential $\phi$ (which is dimensionless in natural units, so that the condition of weakness is simply $\phi\ll 1$) must be proportional to the potential $\phi$ and to the relative deviation between the couplings with gravity of different neutrino eigenstates, which we denote by $\gamma$: by dimensional arguments, this frequency in natural units is $\omega\sim \gamma\phi E$, where $E$ is the neutrino energy. From this argument, we deduce that the effects of VEP, if present, would be increasingly relevant as the neutrino energy increases.

For this reason, the present and upcoming detectors of high energy neutrinos are promising sources of information for constraining VEP in the neutrino sector. Here we investigate two complementary ways of testing VEP effects, using both atmospheric and astrophysical neutrinos. Atmospheric neutrinos, produced at the top of the atmosphere, can be detected at the IceCube Neutrino Observatory after having propagated for a certain length under an approximately constant gravitational potential: they could therefore be subject to VEP-induced oscillations which can induce visible effects on the flux. We use the non-observation of this signature to impose constraints on the VEP model. Astrophysical neutrinos, on the other hand, could also be subject to VEP-induced oscillations along their propagation paths: we show that the present IceCube data on astrophysical neutrinos are already able to exclude a limited region of the VEP parameter space at 95\% confidence level.

\section{VEP effects on neutrino oscillations} \label{sec:oscillations}

Here we deduce the propagation equation for neutrinos moving in a dense environment of electrons with number density $N_e (\mathbf{r})$ and in the presence of VEP: this is a schematization of the neutrino propagation in Earth, in which the coherent forward scattering on electrons can relevantly change the oscillation properties of neutrinos. Throughout this work we use units in which $\hbar=c=1$.

Let the neutrino state be expressed in the flavor basis as $\ket{\psi}=\sum_{\alpha} c_\alpha \ket{\nu_\alpha}$, with $\alpha=e,\; \mu,\; \tau$. In the presence of a gravitational potential and a VEP, three combinations of flavor eigenstates could be selected to couple diagonally with the gravitational potential with three different couplings. These three combinations would then advance with different phases, leading to a non-standard oscillation pattern. We assume that this new basis of eigenstates $\ket{\tilde{\nu}_a}$ is defined by a new unitary matrix $\tilde{U}$ as
\begin{equation}
    \ket{\nu_\alpha}=\sum_a \tilde{U}_{\alpha a} \ket{\tilde{\nu}_a}.
\end{equation}
Denoting the gravitational potential by $\phi$, these three eigenstates feel a different effective potential $\phi_a=\gamma_a \phi$, where $\gamma_a$ are three dimensionless coefficients: in the standard case, all three of them are equal to 1.

The propagation equations are~\cite{Valdiviesso:2011zz}
\begin{eqnarray}\label{eq:propavep}
i\frac{dc_\alpha}{dl}=\sum_{j,\beta} U_{\beta j} U^*_{\alpha j} \frac{\delta m_j^2}{2E} c_\beta + V(\mathbf{r}) \delta_{\alpha e} c_e
 +2E\phi\sum_{a,\beta} \tilde{U}_{\beta a} \tilde{U}^*_{\alpha a} \gamma_a c_\beta.
\end{eqnarray}

Here $E$ is the neutrino energy, $U_{\alpha j}$ are the components of the PMNS matrix, $\delta m_j$ are the neutrino mass splittings, assuming $\delta m_1^2=0$. Furthermore, $V(\mathbf{r})=\sqrt{2} G_F N_e(\mathbf{r})$ is the matter potential, with $G_F$ being the Fermi constant. 

Since only phase differences can cause measurable changes in the oscillation properties, we can subtract a term $2E\gamma_1 c_\alpha$ from Eq. (\ref{eq:propavep}) and express the result in terms of two dimensionless parameters $\gamma_{21}=\gamma_2-\gamma_1$ and $\gamma_{31}=\gamma_3-\gamma_1$: we will refer to these in general as $\gamma_{ij}$.

The VEP model is therefore parameterized in the end by the two strengths $\gamma_{ij}$ and by the unitary matrix $\tilde{U}$: as a first benchmark choice, we assume $\tilde{U}=U$, and we will later discuss the effects of different choices.

\section{Constraints from atmospheric neutrinos at IceCube} \label{sec:atmoic}

Here we discuss the effects of VEP on the atmospheric neutrinos measured by IceCube. Atmospheric neutrinos are produced by the collision of cosmic-rays in the atmosphere: they subsequently propagate from the top of the atmosphere to the Earth surface. We focus on upgoing neutrinos, produced in the Northern hemisphere, which propagate through the Earth to reach the IceCube experiment in the South Pole, for two motivations: first of all, upgoing neutrinos have only a small contamination by atmospheric muons, which can therefore be neglected. Furthermore, upgoing neutrinos have to traverse a path length of the order of the Earth radius: the longer propagation length, compared to the downgoing neutrinos which are produced in the Southern hemisphere, makes them more sensitive to non-standard oscillation effects. 

The analysis we perform in this section is an effective update of the similar one presented in Ref.~\cite{Esmaili:2014ota}. We focus on the data recently released by the IceCube collaboration~\cite{Abbasi:2021bvk}, using three different data samples: the IC79 data sample after one year of observation, the IC86 data sample in 2011, henceforth denoted as IC86-11, and the IC86 data collected from 2012 to 2018, henceforth denoted as IC86-12/18. All three data samples consist of track events, which allow a rather precise angular reconstruction: indeed in this work we are interested in an angular study of the neutrinos. The reason is that neutrinos at different angles have to traverse different lengths in Earth, and therefore they exhibit different oscillation features.

The three data samples of IC79, IC86-11 and IC86-12/18 are composed respectively of 48362, 61313 and 760923 upgoing neutrino events with a zenith angle between $90^\circ$ and $180^\circ$. In all cases we binned the data in ten bins in $\cos\theta$, where $\theta$ is the zenith angle, from 0 to -1. We adopt the atmospheric neutrino model from Ref.~\cite{Honda:2006qj}, and we determine the expected number of events in each zenith bin in the presence of VEP by numerically solving the transport equations Eq.~\ref{eq:propavep}: we use the preliminary Earth reference model (PREM)~\cite{Dziewonski:1981xy} to model the Earth density. The gravitational potential in which the neutrinos move is dominated by the Great Attractor, resulting in $\phi\sim -3\times 10^{-5}$~\cite{Kenyon:1990ve}: since this value is quite uncertain, we will provide all our results in terms of $\gamma_{ij}\phi$. The effective areas for each data sample are provided together with the corresponding data sample in the recent IceCube data release~\cite{Abbasi:2021bvk}.

For each data sample, we denote the expected number of event in the $i$-th angular bin by $N_i^{\text{theor}}$, depending on $\gamma_{ij}$, and we compare it with the observed number of events $N_i^{\text{data}}$ using the $\chi^2$ function~\cite{Esmaili:2014ota}
\begin{eqnarray}
    \chi^2 (\gamma_{21}\phi, \gamma_{31} \phi,\alpha,\beta)=  \sum_i \frac{[N_i^{\text{data}}-\alpha(1+\beta (0.5+\cos\theta))N_i^{\text{theor}}]^2}{\sigma_{i,\text{stat}}^2+\sigma_{i,\text{sys}}^2} + \frac{(1-\alpha)^2}{\sigma_\alpha^2}+\frac{\beta^2}{\sigma_\beta^2},
\end{eqnarray}
where $\alpha$ and $\beta$ take into account the systematic uncertainties on the normalization and the angular distribution respectively of the atmospheric flux and $N_i^{\text{data}}$ are the number of events per bin; the average values of $\alpha$ and $\beta$ are respectively 1 and 0 and the uncertainties on these values $\sigma_\alpha=0.24$ and $\sigma_\beta=0.04$~\cite{Honda:2006qj}. We consider a statistical uncertainty given by the Poisson estimate $\sigma_{i,\text{stat}}=\sqrt{N_i^{\text{data}}}$ and a systematic uncertainty $\sigma_{i,\text{sys}}=f N_i^{\text{th}}$. Since an independent estimate of the systematic uncertainty is not provided by the IceCube collaboration, we use a value for $f$ which guarantees agreement between the data and the atmospheric fluxes within $90\%$ confidence level, similarly to the approach of Ref. \cite{Esmaili:2014ota}. We treat $\alpha$ and $\beta$ as nuisance parameters and we marginalize over them, obtaining an effectively two-dimensional likelihood.

\begin{figure}
    \centering
    \includegraphics[width=0.45\textwidth]{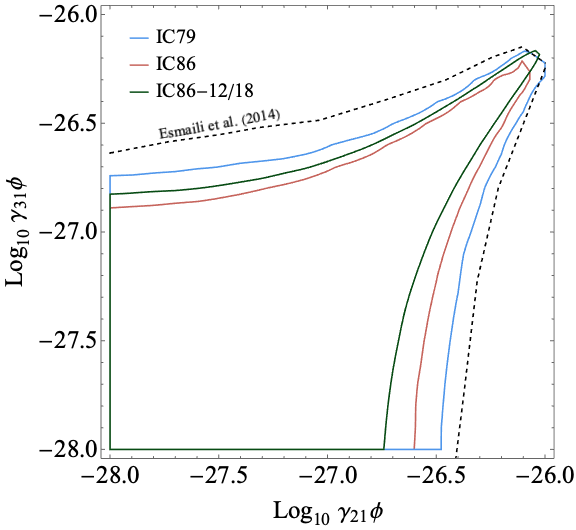}
    \includegraphics[width=0.45\textwidth]{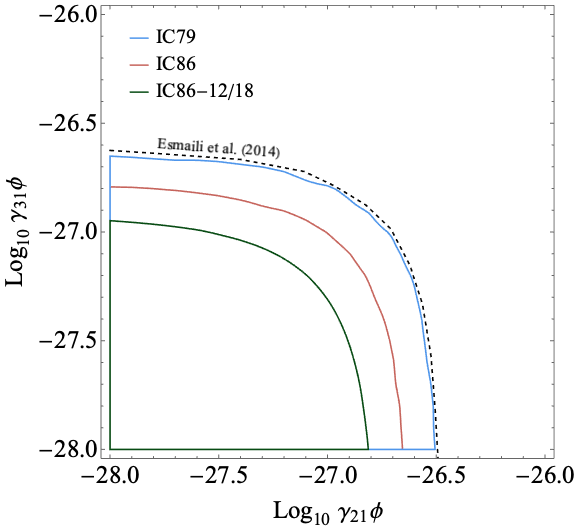}
    \caption{Allowed regions at 90\% confidence level in the $\gamma_{21}-\gamma_{31}$ plane for the IC79, IC86-11 and IC86-12/18 dataset shown in blue, orange and green respectively for the VEP direct (left) and inverse (right) ordering scenario. The black dashed curve is the 90\% confidence level exclusion contour obtained in Ref.~\cite{Esmaili:2014ota}.}
    \label{fig:contatm}
\end{figure}
We show the resulting exclusion contours on the parameters $\gamma_{21}\phi$ and $\gamma_{31}\phi$ in Fig.~\ref{fig:contatm} for two different model choice: in the left panel we show the case of $\gamma_{21}$ and $\gamma_{31}$ having the same signs, while in the right panel we show the case of $\gamma_{21}$ and $\gamma_{31}$ having opposite signs. We also show the previous exclusion contours obtained in Ref.~\cite{Esmaili:2014ota}.

Our results depend on the systematic uncertainty, which here is estimated from the data themselves: an independent estimate of the systematic uncertainty could provide a more robust analysis. Nevertheless, the results clearly show the potential of IceCube in constraining possible VEP.

Up until now we have focused on the choice $\tilde{U}=U$. However, there is no a priori reason why the diagonal basis for coupling with gravity, which we dub gravitational basis, should be connected with the Yukawa eigenstates. The question naturally arises as to how our results change for a different choice. A general analysis spanning the whole parameter space would of course be overwhelming in terms of computational time. In fact, the unitary matrix $\tilde{U}$ is parametrized in terms of three angles and one phase, if neutrinos are of Dirac nature. Here we only limit ourselves to observe that some specific choices could lead to much weaker constraints: an extreme choice is the case in which the gravitational basis coincide with the flavor basis ($\tilde{U}=1$), which would lead to no oscillation at all (at least for energies above 100 GeV, where the vacuum oscillation term does not interfere with the non-standard VEP oscillations) and therefore would cause no observable effects on atmospheric neutrinos.

\section{Effects on astrophysical neutrinos}

In the conclusion to the previous section, we have shown that some choices of the gravitational basis lead to weak or even to no constraint on the VEP model. However, it is still possible to investigate these cases by using the IceCube measurements of astrophysical neutrinos. Indeed astrophysical neutrinos propagating from their sources to the Earth could be subject to non-standard VEP oscillations, leading to a flavor composition which would not be expected in the standard case. The final flavor composition measured at Earth depends on the interplay between standard oscillations and VEP-induced oscillations: since the latter depend on the external gravitational potential, the flavor composition could be rather difficult to predict in general, since the intergalactic gravitational potential is not known with precision.

However, there is a simple scenario which leads to quite definite conclusions independently of the details of the intergalactic gravitational potential. If the VEP is so strong that its effect dominate over those of conventional oscillations over the entire path length, neutrinos will be subject to non-standard oscillations with a wavelength much shorter than the propagation length. In this scenario, the flavor composition at the Earth will be obtained by averaging the probabilities of oscillation over a cycle of oscillation, analogously to what happens in the standard (i.e., no VEP) scenario, where conventional oscillations also have a very short wavelength. However, since in the VEP-dominated scenario the oscillation probabilities are determined by the gravitational basis, and therefore by the choice of $\tilde{U}$, the average oscillation probabilities will be determined by the latter as
\begin{equation}\label{eq:nonstandardaverage}
\tilde{P}_{\alpha\to\beta}=\sum_i |\tilde{U}_{\alpha i}|^2  |\tilde{U}_{\beta i}|^2,
\end{equation}
where $\tilde{P}_{\alpha\to\beta}$ is the probability that an $\alpha$ flavor neutrino produced at the source is converted to a $\beta$ flavor neutrino at the Earth. This relation is to be compared with the oscillation probability in the standard case
\begin{equation}\label{eq:standardaverage}
P_{\alpha\to\beta}=\sum_i |U_{\alpha i}|^2  |U_{\beta i}|^2.
\end{equation}
It is clear that for the specific choice $\tilde{U}=U$, which was used as a benchmark in the previous section, the VEP-dominated scenario will lead to no observable consequences different from the standard case. However, for other choices of the matrix $\tilde{U}$, VEP could cause significant deviations from the standard predictions. Before illustrating these consequences, let us comment on the strength of VEP required to attain such a VEP-dominated scenario. The condition to be fulfilled is that the typical inverse VEP wavelength, of order $\gamma_{ij}\phi E$, is much larger than the inverse wavelength for conventional oscillations, of order $\delta m_i^2 E^{-1}$: for a typical energy larger than about 100 TeV, this condition is fulfilled if $\gamma_{ij}\phi \gtrsim 10^{-31}$. As discussed in Ref. \cite{Minakata:1996nd}, a typical value for the intergalactic potential is $\phi\sim 5\times 10^{-6}$, allowing to test values of $\gamma_{ij}\sim 10^{-26}$.

\begin{figure}
    \centering
    \includegraphics[width=0.7\textwidth]{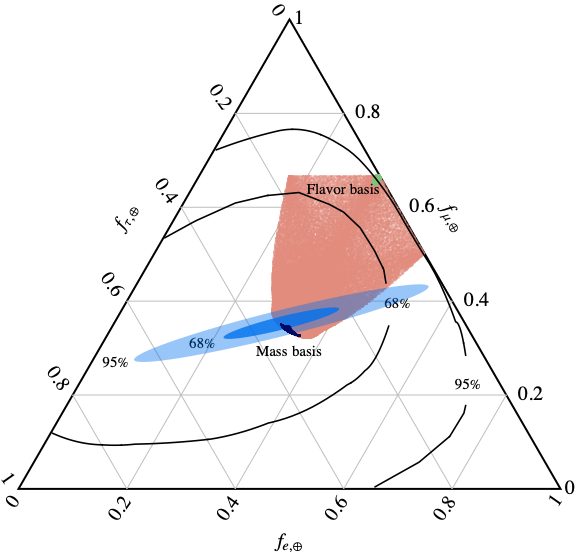}
    \caption{Flavor ratio at the Earth from a pion beam source after averaged VEP oscillations with $\gamma\phi>10^{-31}$: the orange region spans the whole possibilities for the gravitational basis; the green region corresponds to the case of gravitational basis coinciding with flavor eigenstates, while the black region corresponds to the case of gravitational basis coinciding with mass eigenstates. In the latter case we vary the oscillation parameters in the $3$ $\sigma$ intervals. The 68\% and 95\% confidence level obtained by IceCube \cite{Abbasi:2020zmr} are shown (black contours), as well as the projected flavor sensitivity of IceCube-Gen2 \cite{Bustamante:2019sdb} at 68\% and 95\% confidence level (blue ellipses).}
    \label{fig:flavtriang}
\end{figure}

To illustrate the effects that VEP-dominated oscillations could produce on astrophysical neutrinos, we make the simple assumption that astrophysical sources produce neutrinos in the pion beam regime, with a flavor composition at the source $(1:2:0)$. In the standard scenario, the flavor composition at Earth would be obtained using Eq.~\ref{eq:standardaverage}, leading to a flavor composition about $(1:1:1)$. In the VEP-dominated scenario, the oscillation probabilities are given by Eq.~\ref{eq:nonstandardaverage}, and the final flavor composition depends on the matrix $\tilde{U}$. In Fig.~\ref{fig:flavtriang} we show the flavor compositions which can be obtained, starting from a pion beam production, by all possible choices of the unitary matrix for $\tilde{U}$ in the orange region. This region coincides with the region determined in Refs.~\cite{Arguelles:2015dca,Ahlers:2018yom}. Two special choices are given by the case $\tilde{U}=U$ (gravitational basis coinciding with the mass basis), in which the predictions are identical to the standard scenario, and $\tilde{U}$ (gravitational basis coinciding with the flavor basis), in which the flavor composition at Earth remains equal to the flavor composition at the source. We also show the exclusion contours obtained by IceCube. The present information provided by IceCube is not generally able to significantly constrain the VEP parameters, since most of the orange region in Fig. \ref{fig:flavtriang} lies within the 95$\%$ confidence level of the experiment: however, it is noteworthy that a small number of choices, corresponding to the upward right corner of the orange region, are already outside the 95$\%$ contours, meaning that for such choices $\gamma_{ij} \phi>10^{-31}$ is already excluded at $2\sigma$ level. This includes the case in which the gravitational basis coincides with the flavor basis. Interestingly, this is just the case that was not testable by observation of atmospheric neutrinos. We conclude that atmospheric and astrophysical neutrinos are able to test complementary choices for the matrix $\tilde{U}$.

\section{Conclusions}

We have shown that VEP can be tested by neutrino telescopes along two complementary directions: either through the angular and energy distribution of atmospheric neutrinos, or through the flavor composition of astrophysical neutrinos. Using atmospheric neutrinos, we have shown that already with the present data the strength of VEP can be constrained at the level of $\gamma_{ij}\phi\lesssim 10^{-27}$. These two approaches are really complementary in that they can test separate parts of the parameter space: in particular, atmospheric neutrinos are able to test models in which the gravitational basis is substantially different from the flavor basis, so that VEP induces non-standard flavor oscillations; astrophysical neutrinos, on the other hand, are able to test models, in which the gravitational basis is substantially different from the mass basis, so that VEP induces oscillations different from the conventional ones. The combination of the two approaches may lead to a uniform coverage of the parameter space of VEP. 

\acknowledgments{This work was supported by the Italian grant 2017W4HA7S “NAT-NET: Neutrino and As- troparticle Theory Network” (PRIN 2017) funded by the Italian Ministero dell’Istruzione, dell’Universit\'a e della Ricerca (MIUR), and Iniziativa Specifica TAsP of INFN.}


\begin{thebibliography}{99}
\bibitem{Valdiviesso:2011zz}
G.A. Valdiviesso, M.M. Guzzo, and P.C. de~Holanda.
\newblock {Probing new limits for the Violation of the Equivalence Principle in
  the solar\textendash{}reactor neutrino sector as a next to leading order
  effect}.
\newblock {\em Phys. Lett. B}, 701:240--247, 2011.

\bibitem{Esmaili:2014ota}
A.~Esmaili, D.R. Gratieri, M.M. Guzzo, P.C. de~Holanda, O.L.G. Peres, and G.A.
  Valdiviesso.
\newblock {Constraining the violation of the equivalence principle with IceCube
  atmospheric neutrino data}.
\newblock {\em Phys. Rev. D}, 89(11):113003, 2014.

\bibitem{Abbasi:2021bvk}
R.~Abbasi \textit{et al.} [IceCube],
doi:10.21234/CPKQ-K003
[arXiv:2101.09836 [astro-ph.HE]].

\bibitem{Honda:2006qj}
M.~Honda, T.~Kajita, K.~Kasahara, S.~Midorikawa, and T.~Sanuki.
\newblock {Calculation of atmospheric neutrino flux using the interaction model
  calibrated with atmospheric muon data}.
\newblock {\em Phys. Rev. D}, 75:043006, 2007.

\bibitem{Dziewonski:1981xy}
A.M. Dziewonski and D.L. Anderson.
\newblock {Preliminary reference earth model}.
\newblock {\em Phys. Earth Planet. Interiors}, 25:297--356, 1981.

\bibitem{Kenyon:1990ve}
I.R. Kenyon.
\newblock {A Recalculation of the Gravitational Mass Difference Between the
  $K^0$ and $\bar{K}$0 Mesons}.
\newblock {\em Phys. Lett. B}, 237:274--277, 1990.

\bibitem{Minakata:1996nd}
H.~Minakata and A.Yu. Smirnov.
\newblock {High-energy cosmic neutrinos and the equivalence principle}.
\newblock {\em Phys. Rev. D}, 54:3698--3705, 1996.

\bibitem{Abbasi:2020zmr}
R.~Abbasi et~al.
\newblock {Measurement of Astrophysical Tau Neutrinos in IceCube's High-Energy
  Starting Events}.
\newblock 11 2020.

\bibitem{Bustamante:2019sdb}
M.~Bustamante and M.~Ahlers.
\newblock {Inferring the flavor of high-energy astrophysical neutrinos at their
  sources}.
\newblock {\em Phys. Rev. Lett.}, 122(24):241101, 2019.

\bibitem{Arguelles:2015dca}
Carlos~A. Arg\"uelles, Teppei Katori, and Jordi Salvado.
\newblock {New Physics in Astrophysical Neutrino Flavor}.
\newblock {\em Phys. Rev. Lett.}, 115:161303, 2015.

\bibitem{Ahlers:2018yom}
Markus Ahlers, Mauricio Bustamante, and Siqiao Mu.
\newblock {Unitarity Bounds of Astrophysical Neutrinos}.
\newblock {\em Phys. Rev. D}, 98(12):123023, 2018.



\end{thebibliography}
\end{document}